# Transient measurement of phononic states with covariance-based stochastic spectroscopy

G. Sparapassi[1,2], S. M. Cavaletto[3], J. Tollerud[4*], A. Montanaro[1,2], F. Glerean[1,2], A. Marciniak[1,2], F. Giusti[1,2], S. Mukamel[3], D. Fausti[1,2*]

[1] Physics Department, University of Trieste, 34127 Trieste, Italy;
[2] Elettra-Sincrotrone Trieste S.C.p.A., 34149 Trieste, Italy;
[3] Department of Chemistry and Department of Physics & Astronomy, University of California, Irvine, CA 92697, USA;
[4] Optical Sciences Centre, Swinburne University, Melbourne, Australia, 3122;
* Correspondence: daniele.fausti@elettra.eu; jtollerud@swin.edu.au

Abstract:

We present a novel approach to transient Raman spectroscopy, which combines stochastic probe pulses and a covariance-based detection to measure stimulated Raman signals in alpha-quartz. A coherent broadband pump is used to simultaneously impulsively excite a range of different phonon modes, and the phase, amplitude, and energy of each mode are independently recovered as a function of the pump–probe delay by a noisy-probe and covariance-based analysis. Our experimental results and the associated theoretical description demonstrate the feasibility of 2D-Raman experiments based on the stochastic-probe schemes, with new capabilities not available in equivalent mean-value-based 2D-Raman techniques. This work unlocks the gate for nonlinear spectroscopies to capitalize on the information hidden within the noise and overlooked by a mean-value analysis.

Introduction:

Nonlinear optics represents an active research field, dealing with phenomena occurring when intense light interacts with a material. This field is advanced by frequent technological and theoretical developments, delivering a deeper understanding of the properties of matter and revealing delicate interplays between different degrees of freedom. For instance, time-resolved spectroscopies as Raman scattering techniques employ pairs of ultrashort pulses to track the dynamics of transient photo-induced electronic states [1-3]. Tailored time or wavevector combinations of multiple pulses are employed to extract specific nonlinear responses from a sample in multidimensional spectroscopy, a powerful nonlinear technique exploring the complex energy landscape of biological samples [4,5], quantum wells [6], or polymers[7].

However, nonlinear optical techniques rely on extremely weak signals, often orders of magnitude weaker than the linear response[8]. This challenge has been circumvented largely by creative signal-isolation strategies[9-12] and advancements in mathematical descriptions of nonlinear signals[8,13]. It has also been enabled by extensive effort and investment in stable laser systems and experimental setups[9]. Most experimental techniques rely on this stability and a mean-value analysis framework, in which the signal is measured in an integrated fashion (e.g. at the detector level or through repeated measurements) to reduce the noise until a suitable signal-to-noise ratio is achieved.

This approach has some drawbacks. Higher-order measurements become increasingly challenging due to the rapid diminishment of the signal efficiency, making anything beyond 5[th]-order experimentally impractical in many cases. In certain contexts, this is quite limiting. For example, techniques aiming to measure a Raman-echo (the Raman equivalent of photon-echo[14,15] and spin-echo[16] techniques which have proven extremely effective for multidimensional spectroscopy at IR/optical[4,9,17,18] and radio frequencies[19,20], respectively) attracted significant effort in the 90's and early 2000's but have not seen widespread adoption due to the requirement of 7[th]-order signal[21-23]. 5[th]-order 2D-Raman equivalents have been developed and remain an active area of research even though they only provide a subset of the capabilities of a 7th-order Raman echo technique.[24-29]

The mean-value detection is intrinsically blind to the information contained in the noise. We have previously demonstrated a paradigmatically different framework in which the noise is seen as an asset rather than a liability which accesses information the mean-value misses[30] and similar schemes have also recently been reported[31,32]. In our initial demonstration we resolved the impulsive-stimulated Raman scattering (ISRS) spectrum of quartz using a single beam experiment. The nonlinear interaction of noisy input pulses with the Raman-active modes in the quartz generated spectral correlations that we resolved through a covariance-based analysis technique.

In this work, we apply the same covariance-based analysis framework to a transient ISRS measurement. Unlike traditional mean-value-based ISRS measurements[33,34], the covariance-based framework allows us to resolve the phase, amplitude, and frequency of each mode excited impulsively by the broadband coherent pump pulse for each pump-probe delay. We can thus clearly separate contributions from different modes and identify signals even when many modes are simultaneously excited by the pump.

The measurement is well described by a 3rd-order time-resolved ISRS model, where we fully account for the correlation properties of the stochastic probe pulse used. The model recovers the experimental features observed, and fully supports our interpretation of the signal in this novel configuration.

In addition to improving upon mean-value-based ISRS techniques, this correlation-based experimental framework could be extended to 5th-order by adding a second pump pulse to perform a 2D Raman measurement. Compared with current techniques, this would provide additional insight by spectrally resolving the energy of the final interaction[28,34,35] and could be used to measure spectral diffusion and population dynamics which in a mean-value-based framework require a 7th-order experiment. Our proof-of-principle results and theory supporting them demonstrate the feasibility of such a measurement.

Experiment and results:

In a previous single-beam experiment we showed that a broadband pulse with a stochastic spectral phase can resolve Raman spectra through a correlation-based analysis of pulse-to-pulse intensity fluctuations[30]. In this framework, the nonlinear response of the sample is imprinted in the intensity distribution, in the form of a statistical correlation between different frequencies in the transmitted pulse spectrum. A correlation develops when the frequency difference of two spectral components within the pulse spectral bandwidth matches the low-energy vibrational levels of the crystal. A Raman spectrum can thus be extracted from the statistical distribution of the frequency-resolved intensity.

Here, we combine the stochastic probe with a spectrally coherent pump pulse to study transient ISRS in α-quartz with a correlation-based measurement as depicted in Fig. 1a. As shown in Fig. 1b, the probe pulse is randomized on a single-shot basis using a diffraction-based pulse shaper, and the transmitted spectrum of thousands of unique pulses is acquired for each pump–probe time delay $\Delta t$. Mean-value detection completely overlooks the information in the probe pulse, because the stochastic transmitted spectra average out to zero. To retrieve information, we thus exploit a covariance-based analysis and calculate the Pearson correlation (the covariance over multiple repeated measurements divided by the standard deviation) of the spectrally resolved intensity of the transmitted pulse $I_{OUT}(\omega_{OUT})$ with a reference pulse $I_{IN}(\omega_{IN})$ routed around the sample. In this context, the Pearson coefficient quantifies the degree to which a stochastic intensity fluctuation at $\omega_{IN}$ induces a separate intensity fluctuation at $\omega_{OUT}$ through the nonlinear ISRS interaction of the light with the sample. This calculation is repeated for all combinations of $\omega_{OUT}$ and $\omega_{IN}$, resulting in a three-dimensional matrix $\rho_c(\omega_{IN}, \omega_{OUT}, \Delta t)$, from which we can extract spectrally and

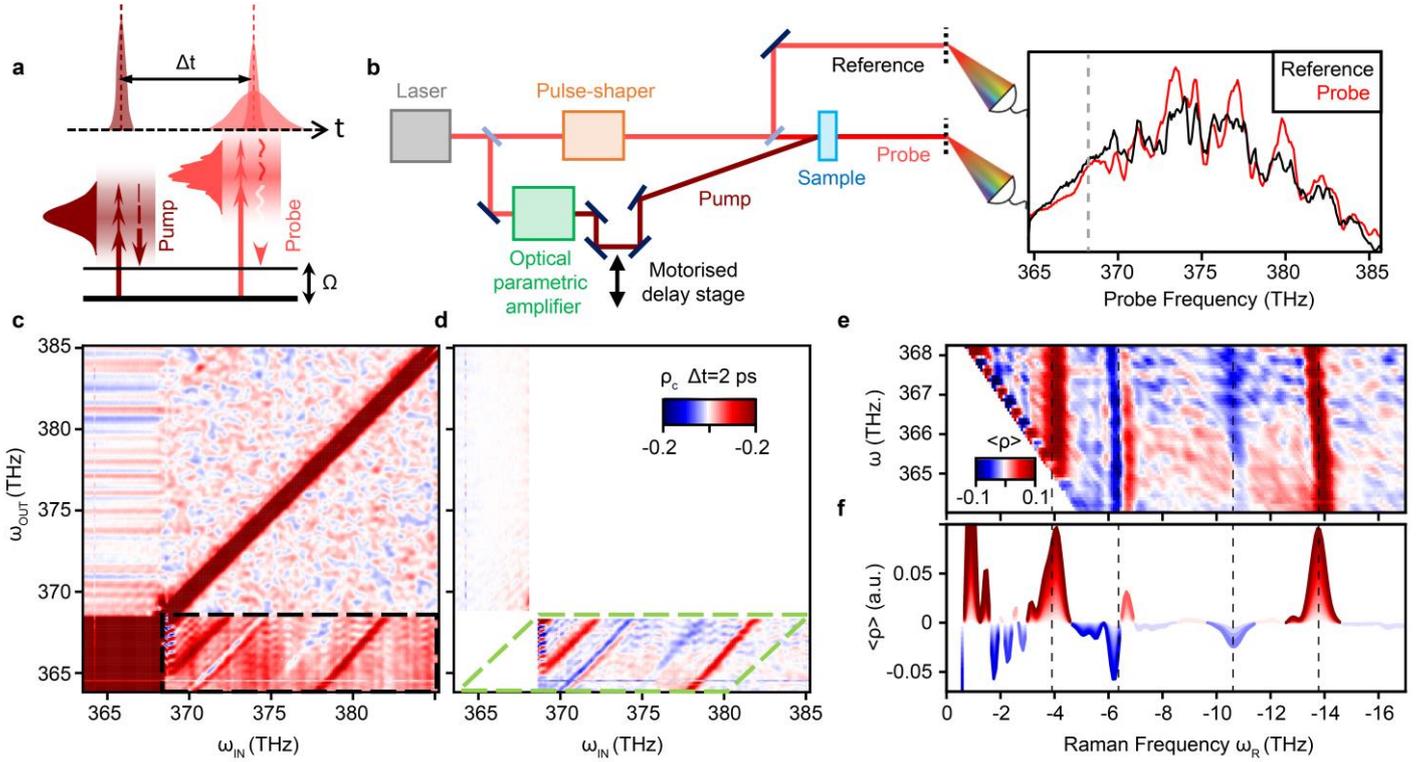

**FIGURE 1 a)** Diagram of the ISRS interaction. **b)** Time-domain representation of the experiment. Spectra of the probe (red) and reference (black) are recorded at each time delay $\Delta t$ and used to calculate the spectrally resolved correlation coefficient $\rho_c$. The vertical grey line indicates the cut-off frequency above which uncorrelated noise was introduced. Pseudo-colormaps of $\rho_c$ for $\Delta t$ = 2 ps **c)** as measured and **d)** after subtracting the median in the lower-right and upper-left quadrant for each $\omega_{IN}$ and $\omega_{OUT}$ (respectively). **e)** The block within the dashed green line in d) is recast by shifting each row from d) by $\omega_{OUT}$ and plotted as a function of the frequency difference $\omega_R$. **f)** The matrix in e) is averaged over $\omega_{OUT}$ yielding $<\rho_c(\omega_R)>$. Black dashed vertical lines indicate the frequency of known Raman modes of quartz.

temporally resolved information. The correlation properties of the stochastic probe pulses used in the experiment are described in the SI. The time profile of a randomized pulse contains a strong central spike, whose width is comparable to the pristine laser pulse duration, and a noisy tail contributed by the added spectral noise. The short coherent spike dictates the time resolution of the experiment, which can resolve the coherent evolution of the phonon even with long noisy tails (> 1ps). We introduce the stochastic phase on the high-energy side of the pulse (>368THz), while leaving lower frequencies noise free, providing a reference field for self-heterodyne measurement of the signal. We will focus on the lower right quadrant (indicated by the dashed black box in Fig. 1c) which is most sensitive to signals that appear for $\omega_{OUT}$ and $\omega_{IN}$ below and above 368 THz (respectively).

In this quadrant the transmitted probe is stochastic, but the reference is not and the ISRS signal produces a uniform diagonal feature for each mode that is shifted away from the $\omega_{IN} = \omega_{OUT}$ diagonal by the phonon frequency. We thus introduce the Raman frequency $\omega_R = \omega_{OUT} - \omega_{IN}$, redefine our frequency axes as $\omega_{OUT}$ and $\omega_R$, and then transform the data onto this new grid by shifting each row by $\omega_{OUT}$ such that the diagonal features of unitary slope become vertical lines. We then project the resulting 2D array onto the $\omega_R$ axis by integrating across $\omega_{OUT}$ (as shown in Figure 1c-f). This compresses each correlation map into a single one-dimensional array which can be conveniently plotted and analysed as a function of the pump and probe delay ($\Delta t$). The result is a two-dimensional map $<\rho_c(\omega_R, \Delta t)>$.

$\rho_c(\omega_{IN}, \omega_{OUT})$ maps for selected delays are shown in Figure 2a. The ISRS signal emerges after the pump excitation as diagonal stripes of nonzero correlation, whose sign evolves with the time delay. Each map is integrated as described above, to produce a single one-dimensional array for each time delay (see Figure 2b)

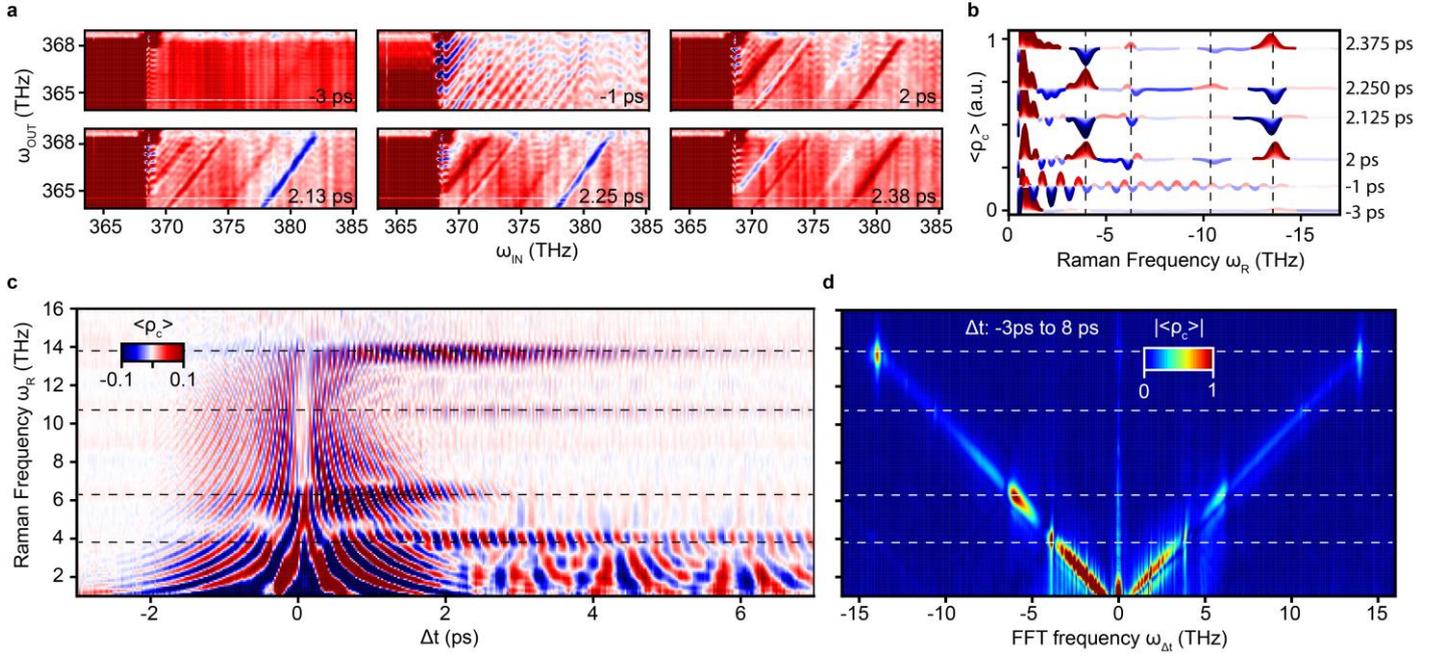

**FIGURE 2** a) $\rho_c$ ($\omega_{IN}$, $\omega_{OUT}$) and b) <$\rho_c(\omega_R)$> for selected values of the time delay $\Delta t$, showing the sign evolution of the vibrational signal. c) Pseudo-colormap of <$\rho_c$ ($\omega_R$, $\Delta t$)>. d) 2D ISRS spectrum, achieved by Hilbert, and Fourier transforms of <$\rho_c$ ($\omega_R$, $\Delta t$)> as a function of $\omega_R$ and $\Delta t$, respectively. Vertical (horizontal) dashed lines in b. (c. and d.) indicate the frequency of known Raman modes of quartz.

and stacked horizontally to build the frequency vs. time $<\rho_c(\omega_R, \Delta t)>$ map (see Figure 2c), that shows the time evolution of the multimode correlation signal. The ISRS signal thus appears as a narrow feature in $\omega_R$ that oscillates as a function of $\Delta t$, and which overlaps a "coherent artefact"-like signal[36,37] at short times (the coherent artefact is described below).

A Hilbert transformation is applied as a function of $\omega_R$ to rotate the phase of the real-valued $<\rho_c(\omega_R, \Delta t)>$ yielding a complex-valued dataset. A Fourier transform is then applied as a function of $\Delta t$, resulting in a frequency–frequency map (see Figure 2d) with the new axis $\omega_{\Delta t}$. In the $<\rho_c(\omega_R, \omega_{\Delta t})>$ map (referred to as a 2D spectrum), we find the phonon spectrum along the vertical axis, and the frequency of oscillations along the horizontal axis. The probe interaction with the active vibrations gives rise to peaks in the 2D spectrum, located at $\omega_{\Delta t} = \pm\omega_R = \pm\Omega$ for each mode $\Omega$ of α-quartz. The frequency of known vibrational modes in quartz are indicated by the horizontal dashed lines. This 2D spectrum is obtained scanning only one inter-pulse delay and exploiting the correlation over the broad bandwidth of the randomized probe pulse to monitor the Raman mode directly. The dephasing rate and the energy of the modes are apparent in the FFT width and position, respectively.

When the pulses temporally overlap ($|\Delta t| < 1.5$ ps), the pulse ordering is undetermined and additional signals overlap the ISRS signal of interest. This effect (often referred to as a coherent artefact or cross-phase modulation) results in a symmetrical signal along $\Delta t$[38] (as can be seen in Fig. 2c) made up of fringes across the spectrum whose periodicity decreases with an increasing $|\Delta t|$. The resulting correlation is an oscillating background, whose fronts are parallel to the correlation-map diagonal (as can be seen in Fig. 2a, -1 ps). In the $<\rho_c(\omega_R, \Delta t)>$ map, this overlap signal resembles a hyperbola, centred at the axes' origin (as can be seen in Fig. 2c). In the 2D spectrum $<\rho_c(\omega_R, \omega_{\Delta t})>$, this is a diagonal feature along the $\omega_R = \pm\omega_{\Delta t}$ lines (as can be seen in Fig. 2d).

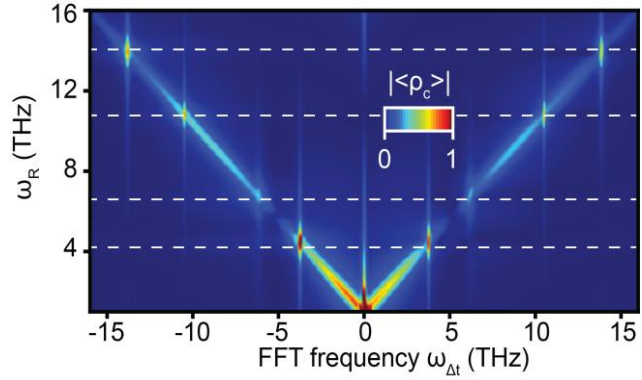

**Figure 3** Simulated ISRS spectrum using the model described in the main text and Supplementary Information. Horizontal lines indicate the frequency of known Raman modes of quartz.

The experiment is well described by a 3$^{rd}$-order model reported in detail in the SI. The double manifold representing the sample is coupled by the pulse electric fields inducing an electronically off-resonant Raman process[13]. Pairs of interactions within the pump spectrum, at frequencies whose difference matches the vibrational energy, create a coherent superposition of vibrational states. The coherence evolves over the pump–probe time delay Δt and is monitored by a subsequent off-resonant Raman excitation involving two interactions with the stochastic probe pulse. The resulting ISRS signal, obtained by frequency dispersing the transmitted probe pulse at varying pump–probe time delays, features Stokes- and anti-Stokes-type contributions, with time-delay-dependent oscillations resulting from the phase of the coherence encountered by the probe pulse[39,40]. The covariance signal, given by products of IN and OUT pulse intensities and thus involving four interactions with the stochastic probe pulse, is calculated in terms of the two- and four-point field correlation functions, taking into account the peculiar stochastic properties of the pulses generated in the experiment and the impact of the gating performed by the discrete detector array.

The calculated 2D ISRS spectrum for the phonon frequencies involved in the experiment is shown in Fig. 3. Here, the diagonal contributions stemming from overlapping pump and probe pulses are added to the ISRS signal. The measured spectrum in Fig. 2d is mostly reproduced by the model as shown in Fig. 3. However, the different peak amplitude of the 6 THz phonon in the two quadrants of the experimental 2D ISRS spectrum is not captured in the 3$^{rd}$-order model, pointing at higher-order contributions, such as a phonon-phonon coupling[41].

Phonon dephasing times can be extracted by fitting the phonon linewidth in the $\omega_{\Delta t}$ domain or by fitting the exponential decay in the Δt domain. We fit the three most prominent phonon features (3.9 THz, 6.2 THz, 13.9 THz) in Δt domain and compared the dephasing times extracted using traditional mean-value-based pump-probe spectroscopy (see SI for details). We found that the dephasing times measured using the covariance-based and mean-value-based approaches were in good agreement.

One challenge with this technique is to separate the ISRS from the coherent-artefact signal, which overlaps the Raman modes in the 2D spectrum and partially in $< \rho_c(\omega_R, \Delta t) >$. One way to separate these two contributions is by applying a window function in the time domain. This can be seen in Fig. 4, where multiple window functions have been applied to separate the different contributions. Clearly, removing the entire 2 ps overlap is not suitable for modes with short dephasing times (e.g. the 6 THz mode). The duration of the noisy tails in the probe pulse is inversely proportional to the correlation length of the stochastic probe phase, so the length of the pump–probe overlap signal can also be reduced, but at the cost of diminished spectral resolution. In practice, an optimal correlation length would be chosen by balancing these two factors based on the needs of the experiment.

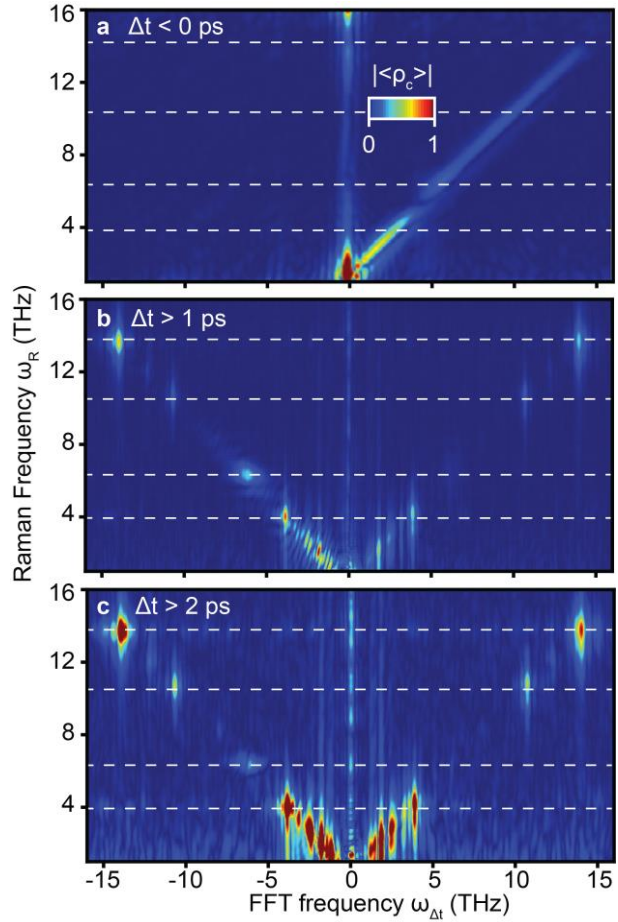

**FIGURE 4** Different contributions to the ISRS signal can be resolved depending on how the data is windowed in the time domain, Δt. a) When selecting the negative time delays, the only contribution to the 2D spectrum is the 1:1 diagonal. When the pump-probe overlap is b) partially and c) fully excluded, the overlap diagonals are reduced and fully removed (respectively), leaving only discrete ISRS peaks. Horizontal lines indicate the frequency of known Raman modes of quartz.

The spectrally uncorrelated fluctuations in the probe pulses are critical to the functionality of this experiment, but uncorrelated noise from other sources has a detrimental effect and may hide the presence of the correlations induced by the sample. The primary source of this unwanted spectrally uncorrelated noise is in the detection system, which generates a 'white' noise spread across the covariance maps. Through an analysis provided in the SI, we show that this uncorrelated noise is at least an order of magnitude below the peak signal for most of the observed phonon modes, and can be reduced by increasing the number of unique phase patterns applied to the probe pulse for each time delay.

Discussion:

The ability to resolve the energy of the Raman modes is undoubtedly a strength of this correlation-based framework compared with mean-value ISRS, but the ability to track the phase of the mode as a function of the pump–probe delay is also useful. First, this phase evolution provides a "sanity-check", as the frequency of the mode generated in the final state must match the frequency of the oscillations in the pump-probe time delay, but it also enables separating the signal along two different frequency axes, so overlapping signals (such as closely spaced modes) are more readily distinguished. This can be seen especially clearly in the low-energy modes in Fig. 4c, which cannot be easily separated in the correlation maps alone.

We observe different amplitudes of the Raman peaks for positive and negative $\omega_{\Delta t}$, even when taking into account the overlap signal. This is in contrast with the theory, which predicts symmetric spectral amplitudes.

The most apparent mismatch is found in correspondence of the combination signals involving the 6 THz phonon. This discrepancy could be due to a higher-order coupling to other elementary excitations in the sample, possibly a beating resulting from an anharmonic interaction[42-44], or to a nontrivial decay channel for the aforementioned phonon[45], which is the widest (and one of the most intense) of those within the probe spectral bandwidth.

It is well known that a 3$^{rd}$-order measurement does not contain information on the pure decoherence of the vibrational modes, and that a 7$^{th}$-order measurement (with four pairs of light–matter interactions) is required to measure the Raman equivalent of 2D-IR photon-echo experiments (typically referred to as a Raman-echo). The success of 2D-IR (and analogues in RF and the visible) suggests that such a Raman-echo experiment would provide very useful information, but it is experimentally challenging to the point of impracticality. As such, 5$^{th}$-order Raman experiments (typically referred to as 2D-Raman) have been developed and can access some of the same information, but importantly, they lack sensitivity to population dynamics that have made the equivalent 2D-IR techniques so effective[28,35].

A 5$^{th}$-order 2D-Raman measurement with covariance-based detection would have some key advantages over mean-value-based techniques. It is intrinsically sensitive to weak signals, because even a small signal can generate a strong correlation. Being able to resolve the phase and energy of the final interaction will enable new ways to separate 5$^{th}$-order signals from 3$^{rd}$-order cascades, which is a common challenge in 2D Raman[46,47]. Finally, we expect that it will enable fast measurement of 2D spectra because only a single delay needs to be scanned, unlike current techniques which require scanning of two delays.

More importantly, 5$^{th}$-order 2D-Raman with covariance-based detection will enable measurement of some population signals, which is not possible in current mean-value based techniques. The capability of the covariance-based detection to resolve the energy of the final state is functionally analogous (albeit imperfectly) to the role of spectrally-resolved or time-resolved heterodyne detection in 2D-IR and 2D electronic spectroscopy, which are phase-resolved photon echo techniques requiring three characteristic time domains[9,10,20]. The state of the system during two of the time domains is determined by scanning the two inter-pulse delays while tracking the phase of the signal, but the state of the system in the third time period after the final interaction (which is critical for capturing the photon echo) is determined by interference between the signal and a reference field. Without this heterodyne detection scheme, a fourth excitation pulse is needed to probe the system and resolve the dynamics in this final time delay[10].

In a Raman-echo measurement, each of these interactions is doubled, so four pairs of interactions are required to generate the three characteristic time periods[34]. However, the covariance-based detection allows us to infer the state of our system in the final time period through the correlations induced in the probe pulse by the final Raman interaction. We can then generate a non-rephasing 2D-spectrum in which one axis is the inferred state of the system in the final time domain (measured through the spectral correlations in the probe) and the other is the state of the system in the first time domain (the phase evolution of the spectral correlations as a function of the inter-pulse delay). The population dynamics of the non-rephasing signals could then be probed by collecting 2D spectra as a function of the second time delay. Importantly, there is no access to rephasing pathways (and thus no separation of homogeneous and inhomogeneous broadening) because the inference will only provide the state of the system immediately after the final interaction, whereas the Raman echo will form subsequently in inhomogeneously broadened modes. Still, this access to the population dynamics would be a powerful new additional capability not possible in current 2D Raman techniques.

Conclusions:

We have shown that the signal from low-energy excitations of a crystalline sample can be mapped within a time-resolved spectroscopic experiment by use of a single delay line in conjunction with a spectrally randomized probe. In this framework, the energy, phase, and amplitude of each Raman mode can be read out directly through correlations in the spectrum of the optical probe pulse. The covariance-based signal measured is consistent with a theoretical model predicting that the oscillation frequency of the signals as a function of the pump–probe delay should match the frequencies of the signals in covariance-based analysis of the probe spectra.

This work is a key step towards covariance-based 2D Raman spectroscopy—an alternative approach to 2D Raman using mean-value signal detection—which promises to decrease experimental complexity, reduce acquisition times, and enable new insights such as measurement of Raman coherence and population dynamics in a $5^{th}$-order experiment. More broadly, this work demonstrates the feasibility of a covariance-based framework in transient nonlinear spectroscopy and its advantages over mean-value-based approaches and will inspire similar efforts to utilize higher modes beyond the mean-value in other nonlinear techniques.

Our work demonstrates that covariance-based spectroscopy is a feasible and advantageous route to retrieve temporal and spectral resolution from stochastic probes. This is also relevant at hard-x-ray frequencies, where current x-ray free-electron lasers (FELs) based on the self-amplified spontaneous emission (SASE) mechanism generate intense pulses which are intrinsically stochastic[48-53]. Covariance-based signals have been used in theory and experimental investigations with stochastic FEL pulses , and were recently shown to provide the same temporal and spectral resolution as signals obtained by coherent pulses[54]. They thus represent an essential ingredient for the implementation of multidimensional nonlinear x-ray spectroscopy[55] with existing technology.

Acknowledgements:

This work was supported by the European Commission through the European Research Council (ERC) Starting Grant Inhomogenieties and Fluctuations in Quantum Coherent Matter Phases by Ultrafast Optical Tomography (INCEPT) (Grant#677488). This work has been performed using the LEGEND laser source made available by the Nanoscience Foundry and Fine Analysis (NFFA-MIUR Italy Progetti Internazionali) facility. S.M.C. gratefully acknowledges the support of the Alexander von Humboldt foundation through the Feodor Lynen program. S.M.C. and S.M. gratefully acknowledge the support of the National Science Foundation (Grant CHE-1953045).

## Methods

The sample is an α-quartz crystal, with 1 mm thickness.

The laser employed in the experiments is a Coherent Legend Elite Duo, producing pulses at a repetition rate of 5 kHz, with wavelength λ = 795 nm, and duration 45 fs. The pump is collected with a window before the beam is sent through the pulse shaper. The wavelength of the pump is down-converted to λ = 1300 nm using a home-made double stage optical parametric amplifier. The pump has a duration of 110 fs (as measured with a homemade FROG system), and a fluence of 4.5 mJ/cm$^2$. The probe is routed into a diffraction-based liquid crystal spatial light modulator [Vaughan2005], that randomizes shot by shot its spectral phase. The resulting average probe pulse, in the time domain, is made of a central spike whose width is essentially untouched by the randomization, and a broad shoulder lasting roughly 1-1.5 ps.

The randomized beam is finally separated into a reference, selected before the sample, and a probe, transmitted by the sample, with 0.025 mJ/cm$^2$ fluence. The probe power is kept low enough that no features can be resolved in correlation maps in which the pump is blocked, so the detected signal is not contaminated by probe-only correlations. While this signal could also in principle result from scatter interference between pump and probe, we rule this out because of the wavelength difference.

A homemade detection system was used. It is composed of a pair of twin Hamamatsu photodiode arrays, triggered by the laser and at the same repetition rate, that digitize the dispersed probe and reference spectra. The actual experimental repetition rate is lowered by the slow rotation dynamics of the liquid crystals the pulse shaper is based on. The frequency resolution, roughly 0.2 THz, is determined by the narrow average width of the spiky spectral profile, in turn given by the random spectral phase correlation length.

---

We apply the following real-to-complex transformation to the frequency vs. time $<\rho_c(\omega_R, \Delta t)>$ map in order to remove the symmetry between the negative and positive Fourier frequency components in the 2D spectrum $<\rho_c(\omega_R, \omega_{\Delta t})>$. This also allows to distinguish the overlap signal from the Raman signal, as the first is symmetrical around the pump probe overlap along the time delay axis, while the latter is only present at positive time delays.

The data in $<\rho_c(\omega_R, \Delta t)>$ is Fourier transformed along the $\omega_R$ axis to produce a $<\rho_c(\Delta t_R, \Delta t)>$ time vs. time map. Only one half of this map is meaningful as the starting data is real valued. We window out the mirrored half and perform an inverse Fourier Transform along the $\Delta t_R$ axis. The $<\rho_c(\omega_R, \Delta t)>$ map we obtain is complex valued, therefore its Fourier Transform along the true time delay axis is meaningful all along the $\omega_{\Delta t}$ axis, in other words the FT spectrum is not mirrored. Before performing this last FT though, we zero pad the time dependent data in order to smooth its edges and remove any artifacts from the 2D spectrum. The procedure just outlined corresponds to performing the Hilbert transform of the signal.